\documentclass{rspublic}

\usepackage{graphicx,natbib,amssymb,float}
\usepackage{mathrsfs,fancyhdr,lastpage}
\usepackage{amssymb}
\usepackage{amsmath}
\usepackage{tikz}
\usepackage{color}
\usepackage{paralist}
\usepackage[normalem]{ulem}
\usetikzlibrary{decorations.markings}
\usetikzlibrary{decorations.pathmorphing}
\usetikzlibrary{fit}
\usetikzlibrary{calc,arrows,patterns}
\usepgflibrary{plotmarks}

\newcommand{\bg}[1]{\mbox{\boldmath $#1$}}

\newcommand{\tr}{\mathrm{tr}\,}

\newcommand{\for}{\mathrm{for}}

\newcommand{\chain}{\mathrm{chain}}

\newcommand{\M}{\mathrm{max}}

\newcommand{\ti}{\mathrm{ti}}
\newcommand{\iso}{\mathrm{i}}
\newcommand{\T}{\mathrm{T}}
\newcommand{\tanhs}{\mathrm{\tanh\,}}

\begin{document}

\title{Transversely isotropic cyclic stress-softening model for the Mullins effect}

\author [S. R. Rickaby and N. H. Scott]{Stephen R. Rickaby\footnote{Email: stephen.r.rickaby@gmail.com} and Nigel H. Scott\footnote{Email: n.scott@uea.ac.uk}}
\affiliation{School of Mathematics, University of East Anglia, Norwich Research Park, Norwich NR4 7TJ, UK}

\maketitle

\begin{center}
\emph{[Received 2 August 2012; Accepted 4 September 2012]}
\end{center}

\begin{abstract}{Mullins effect, stress-softening, hysteresis, stress relaxation, residual strain, creep of residual strain,  transverse 
isotropy.\\
\textbf{MSC codes:}  74B20 $\cdot$ 74D10 $\cdot$ 74L15}
This paper models  stress softening  during cyclic loading and unloading of an elastomer. The paper begins by remodelling the primary loading curve to include a softening function and goes on to derive non-linear transversely isotropic constitutive equations for the elastic response, stress relaxation, residual strain and creep of residual strain.   These ideas are combined with a transversely isotropic version of the Arruda-Boyce eight-chain model to develop a constitutive relation that is capable of accurately representing the Mullins effect during cyclic stress-softening for a transversely isotropic, hyperelastic material, in particular a carbon-filled rubber vulcanizate. 
\end{abstract}

\section{Introduction} 

\thispagestyle{fancy} \lhead{\emph{Proc. R. Soc. A}   (2012) {\bf 468},  4041--4057   \\   
doi:10.1098/rspa.2012.0461\\ 
\emph{Published online 3 October 2012}\\
arXiv:
}
\chead{18 May 2020}
\rhead{Page\  \thepage\ of\ \pageref{LastPage}}
\cfoot{}

When a rubber specimen is loaded, unloaded and then reloaded, the subsequent load required to produce the same deformation is smaller than that required during primary loading. This stress-softening phenomenon is known as the Mullins effect,  named after \cite{mullins1947}  who  conducted an extensive study of carbon filled rubber vulcanizates. \cite{dianib} have written a  recent review of this effect, detailing specific features associated with stress-softening and providing a pr\'ecis of models developed to represent this effect.

Many authors have modelled the Mullins effect since Mullins,  for example, \cite{ogden}, \cite{dorfmann}, \cite{dianib}  and  \cite{tommasi} who present an interesting micromechanical model.  However, most  authors  model a simplified version of the Mullins effect,   neglecting the following inelastic features: hysteresis, stress relaxation, residual strain and creep of residual strain.

\cite{mullins1947} observed experimentally that when a rubber vulcanizate sheet undergoes an equibiaxial tension, softening occurs in all three directions. The degree of softening is not the same in all three directions  and therefore anisotropic stress-strain properties are developed. We expect that any model  capable of representing accurately  the  experimental data on stress-softening would need to take this feature into consideration.

Not all inelastic features may be relevant for a particular application.  Therefore, in order to develop a functional model we require that specific parameters could be set to zero to exclude any particular inelastic feature yet still maintain the integrity of the model.

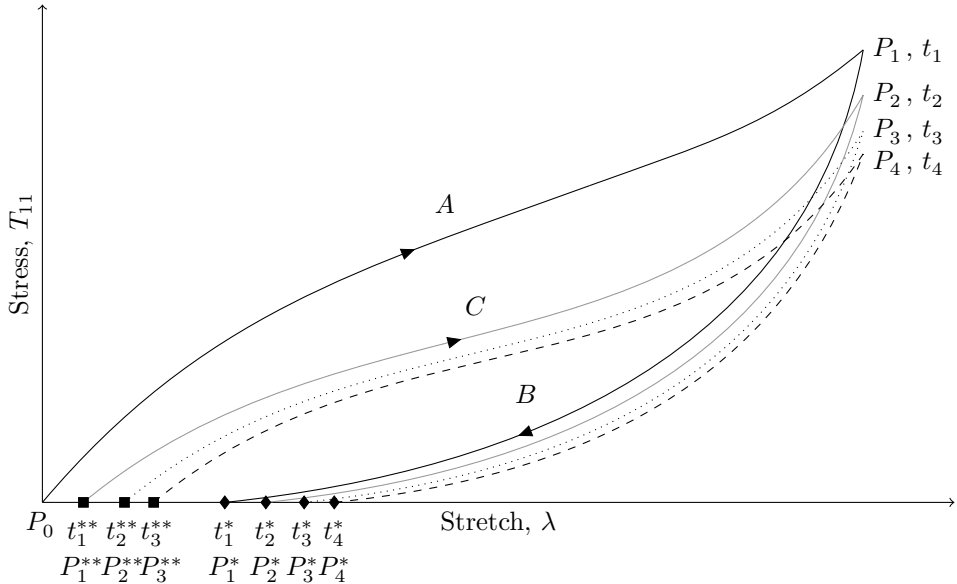
\begin{figure}[h]
\centering
\begin{tikzpicture}[scale=1.20, decoration={
markings,
mark=at position 6cm with {\arrow[black]{triangle 45};},
mark=at position 16.5cm with {\arrowreversed[black]{triangle 45};},
mark=at position 41cm with {\arrow[black]{triangle 45};},}
]
\draw[->] (0,0) -- (10,0)node[sloped,below,midway] {Stretch, $\lambda$} ;
\draw[->] (0,0) -- (0,5.5) node[sloped,above,midway] {Stress, $T_{11}$};
\draw [black](0,0) to[out=50,in=200] node [sloped,above] {} (6,3.5);
\draw [black](6,3.5) to[out=20,in=220] node [sloped,above] {} (9,5);
\draw [black] (2,0) to [out=6,in=260] node [sloped,above] {} (9,5);
\draw [black!40] 
(0.45,0) to[out=40,in=240] node [sloped,above] {} (9,4.5)
(2.45,0) to[out=6,in=256.5] node [sloped,below] {} (9,4.5);
\draw [dotted]
(0.9,0) to[out=40,in=238] node [sloped,below] {} (9,4.1)
(2.87,0) to[out=6,in=253] node [sloped,above] {} (9,4.1);
\draw [dashed] 
(1.22,0) to[out=40,in=236] node [sloped,below] {} (9,3.85)
(3.2,0) to[out=6,in=250] node [sloped,above] {} (9,3.85);
\draw [postaction={decorate}][loosely dotted,line width=0.01pt]
(0,0) to[out=50,in=200] node [sloped,above] {} (6,3.5)
(6,3.5) to[out=20,in=220] node [sloped,above] {} (9,5)
	(2,0) to[out=6,in=260] node [sloped,above] {} (9,5)
	(0.45,0) to[out=40,in=240] node [sloped,above] {} (9,4.5)
	(0.45,0) to[out=40,in=240] node [sloped,above] {} (9,4.5);
\coordinate [label=right:{$P_1^{\phantom{*}}, \,t_1^{\phantom{*}}$}] (B) at (9.0	,5);
\coordinate [label=right:{$P_2^{\phantom{*}}, \,t_2^{\phantom{*}}$}] (B) at (9.0	,4.5);
\coordinate [label=right:{$P_3^{\phantom{*}}, \,t_3^{\phantom{*}}$}] (B) at (9.0	,4.1);
\coordinate [label=right:{$P_4^{\phantom{*}}, \,t_4^{\phantom{*}}$}] (B) at (9.0	,3.75);
\coordinate [label=below:{$t_1^*$}] (B) at (2,-0.1);
\coordinate [label=below:{$t_2^*$}] (B) at (2.45,-0.1);
\coordinate [label=below:{$t_3^*$}] (B) at (2.85,-0.1);
\coordinate [label=below:{$t_4^*$}] (B) at (3.2,-0.1);
\coordinate [label=below:{$t_1^{**}$}] (B) at (0.45,-0.1);
\coordinate [label=below:{$t_2^{**}$}] (B) at (0.88,-0.1);
\coordinate [label=below:{$t_3^{**}$}] (B) at (1.26,-0.1);
\coordinate [label=below:{$P_1^*$}] (B) at (2,-0.5);
\coordinate [label=below:{$P_2^*$}] (B) at (2.45,-0.5);
\coordinate [label=below:{$P_3^*$}] (B) at (2.85,-0.5);
\coordinate [label=below:{$P_4^*$}] (B) at (3.2,-0.5);

\coordinate [label=below:{$P_1^{**}$}] (B) at (0.43,-0.5);
\coordinate [label=below:{$P_2^{**}$}] (B) at (0.88,-0.5);
\coordinate [label=below:{$P_3^{**}$}] (B) at (1.3,-0.5);
\coordinate [label=right:{$A$}] (B) at (4.2,3.3);
\coordinate [label=below:{$B$}] (B) at (5.3,1.4);
\coordinate [label=below:{$C$}] (B) at (4.75,2.37);
\coordinate [label=below:{$ $}] (B) at (0.9,0);
\coordinate [label=below:{$ $}] (B) at (1.2,0);
\coordinate [label=below:{$ $}] (B) at (1.5,0);
\coordinate [label=below:{$P_0^{\phantom{*}}$}] (O) at (0,0);
\draw plot[only marks, mark=diamond*, mark size=2pt] coordinates {(2,0) (2.45,0) (2.87,0) (3.2,0)} ;
\draw plot[only marks, mark=square*, mark size=1.5pt] coordinates {(0.45,0) (0.9,0) (1.22,0) } ;
\end{tikzpicture}
\caption{Cyclic stress-softening with residual strain.}
\label{fig:1}  
\end{figure}

The time dependence of a rubber specimen which is  cyclically stretched up to a particular value of the strain is as represented in Figure \ref{fig:1}. Initially,   loading starts at point $P_0$ at time $t_0$ and the specimen is loaded to the particular strain $\lambda_\M$ at point $P_1$ at  time $t_1$, the material  then being unloaded to zero stress at the point $P_1^*$ at time $t_1^*$ with corresponding strain $\lambda_1^{*}$, where $1<\lambda_1^{*}<\lambda_\M$. Further recovery, known as residual creep,  then  occurs at zero stress before reloading  commences at time $t_1^{**}$ at point $P_1^{**}$ with strain $\lambda_1^{**}$, where $1<\lambda_1^{**}<\lambda_1^{*}$.  This reloading  terminates at the same strain $\lambda_\M$ as before, but now at point $P_2$  and  time $t_2$.  This pattern continues throughout the unloading/reloading process.
 
In this paper we derive a transversely isotropic constitutive model to represent the Mullins effect for cyclic stress-softening under uniaxial tension. In Section \ref{sec:isoelast} we present the isotropic elastic  model as developed by \cite{rickaby}. Section \ref{sec:primary} focuses on developing a stress-softening model for the primary loading path. Section \ref{sec:anisotropic} follows the work of \cite{spencer} and lays the foundations for a transversely isotropic model, which is then developed through Sections \ref{sec:transversely}, \ref{sec:transverselystress} and \ref{sec:residual},  where transversely isotropic models are presented for Arruda-Boyce eight-chain elasticity, stress relaxation and residual creep, respectively.  In  Section \ref{sec:constitutive} we present a new transversely isotropic constitutive model and compare it with  experimental data. Finally, in Section \ref{sec:conclusion} we conclude that the present model of transverse isotropy and the introduction of stress-softening on the primary loading path provide a much better fit to experimental data 
in comparison with the isotropic model of \cite{rickaby}.

Preliminary results of the model were presented in \cite{ricscott}. 

\section{Isotropic elastic response} 
\label{sec:isoelast}

In the reference configuration, at time $t_0$, a material particle is  located at the position $\textbf{X}$ with Cartesian components $X_1,X_2,X_3$ relative to the orthonormal basis $\{\mathbf{e}_1, \mathbf{e}_2, \mathbf{e}_3\}$.   After  deformation, at time $t$, the same particle is located  at the position $\bg{x}(\textbf{X},t)$ with components $x_1,x_2,x_3$  relative to the same orthonormal basis $\{\mathbf{e}_1, \mathbf{e}_2, \mathbf{e}_3\}$.   The deformation gradient is defined by
\begin{equation*}
F_{iA}(\textbf{X},t)=\frac{\partial x_i(\textbf{X},t)}{\partial X_A},\quad{\rm or\;\;simply,}\quad
\textbf{F}(\textbf{X},t)=\frac{\partial \bg{x}(\textbf{X},t)}{\partial \textbf{X}}.
\end{equation*}
An isochoric uniaxial strain is taken in the form
\begin{equation}
x_1=\lambda X_1, \quad x_2=\lambda^{-\frac{1}{2}} X_2, \quad x_3=\lambda^{-\frac{1}{2}} X_3.
\label{eq:2.1z}
\end{equation}
The  right Cauchy-Green strain tensor  
$\textbf{C}=\textbf{F}^\T\textbf{F}$ is given by
\begin{displaymath}
\textbf{C}= \left( \begin{array}{ccc}	
{\lambda^2}& 0 &0\\
0 &{\lambda^{-1}}& 0\\
0& 0 &{\lambda^{-1}} \end{array}\right),
\end{displaymath}
and has 
 principal invariants
\begin{equation}
 I_1 = \tr {\bf C} = \lambda^2+2\lambda^{-1},\quad I_2 = I_3 \,\tr {\bf C}^{-1} = \lambda^{-2}+2\lambda,
\quad I_3 =\det {\bf C}= 1,
\label{eq:2.2z} 
\end{equation}
the last being a consequence of isochoricity.

An incompressible isotropic hyperelastic material possesses a strain energy function $W(I_1, I_2)$ in terms of which the Cauchy stress is given by
\begin{align}
\textbf{T}^{\mathscr{E}_{\iso}}(\lambda) =& \, -p\textbf{I}+2\left[\frac{\partial{W}}{\partial I_1}+I_1\frac{\partial{W}}{\partial I_2}\right]
\textbf{B}-2\frac{\partial{W}}{\partial I_2}\textbf{B}^2,
\label{eq:2.3z}
\end{align}
where $p$ is an arbitrary pressure, $\bf I$ is the unit tensor and $ \textbf{B}= \mathbf{F}\mathbf{F}^\T$ is the left Cauchy-Green strain tensor. 
We are concerned here only with uniaxial tension in the 1-direction and so may fix the value of $p$ by the requirement
\[
\textbf{T}^{\mathscr{E}_{\iso}}_{22}(\lambda)=\textbf{T}^{\mathscr{E}_{\iso}}_{33}(\lambda)=0.
\]
Using this value of $p$ in equation (\ref{eq:2.3z}) then gives the only non-zero component of stress to be the uniaxial tension
\begin{equation}
 T_{11}^{\mathscr{E}_{\iso}}(\lambda) =  2(\lambda^2 - \lambda^{-1})
 \left[\frac{\partial{W}}{\partial I_1}+  \lambda^{-1}\frac{\partial{W}}{\partial I_2}\right],
\label{eq:2.4z}
\end{equation}
in which equation (\ref{eq:2.2z})$_1$ has been used.
The uniaxial tension (\ref{eq:2.4z}) vanishes in the reference configuration, where $\lambda=1$.

The \cite{arruda}  eight-chain model was developed to model non-linear isotropic rubber elasticity by considering the properties of the polymer chains 
of which rubber is composed.  It is characterized by the strain energy function
\begin{equation}
W_{\iso}=\mu N \left\{\sqrt{\frac{I_1}{{3N}}}\,\beta +\log \left(\frac{{\beta}}{\sinh{\beta}}\right) \right\} ,
\label{eq:2.5z}
\end{equation}
where
\begin{equation}
{\beta}=\mathscr{L}^{-1}\left(\sqrt{\frac{I_1}{{3N}}}\right).
\label{eq:2.6z}
\end{equation}
Here, $\mu$ is the shear modulus and $N$ is the number of links forming a single polymer chain.   The  Langevin function is defined by
\[
\mathscr{L}(x)=\coth x-\frac{1}{x}
\]
with  inverse denoted by  $\mathscr{L}^{-1}(x)$.
Upon substituting for $W$ from equation (\ref{eq:2.5z}) into equation (\ref{eq:2.3z}) we obtain the elastic stress in the Arruda-Boyce model:
\begin{align}
\textbf{T}^{\mathscr{E}_{\iso}}(\lambda) =& \, -p \textbf{I}+ \frac{\mu}{3} \sqrt{\frac{3N}{I_1}}\mathscr{L}^{-1}\left(\sqrt{\frac{I_1}{3{N}}}\right)\textbf{B}.
\label{eq:2.7z}
\end{align}

\section{Stress softening} 
\label{sec:primary}
\subsection{Softening on the unloading and reloading paths}
In order to model stress softening on the unloading and reloading paths, i.e. paths $B$ and $C$ of Figure \ref{fig:1},
\cite{rickaby} introduced the following softening function,
\begin{equation}
\zeta_\omega(\lambda) = 1-\frac{1}{r_\omega}\left\{\tanh\left(\frac{W_{\M}-{W}}{\mu b_\omega}\right)\right\}^{{1}/{\vartheta_\omega}}, \label{eq:3.1z}
\end{equation}
where $W_{\M}$ is the maximum strain energy which is achieved on the primary loading path  at the maximum stretch  
$\lambda_{\M}$  before unloading commences.   ${W}$ is the  strain energy value at the intermediate stretch 
$\lambda$, so that $0\leq W\leq W_{\M}$ when $1\leq\lambda\leq\lambda_{\M}$.    The quantities  $b_\omega$, $r_\omega$ are positive dimensionless material constants where
\[
\omega=\left\{ \begin{array}{clrr}
1 & \textrm{unloading}, & \textrm{following path $B$ of  Figure \ref{fig:1}} ,\\[2mm]
2 & \textrm{reloading}, & \textrm{following path $C$ of  Figure \ref{fig:1}}.\\
\end{array}\right.
\]

The softening function  (\ref{eq:3.1z}) has the property that
 \begin{equation}
\textbf{T} =  \zeta_\omega(\lambda) \textbf{T}^{\mathscr{E}_{\iso}}(\lambda),
 \label{eq:3.2z}
 \end{equation}
thus providing a relationship between the Cauchy stress $\textbf{T}$ in the    unloading and reloading of the material, after the primary loading has ceased, and the Cauchy stress  $\textbf{T}^{\mathscr{E}_{\iso}}(\lambda)$ in the primary loading phase of an isotropic elastic parent material.  In the model of \cite{rickaby} the stress 
$\textbf{T}^{\mathscr{E}_{\iso}}(\lambda)$ 
is a purely elastic response, not subject to any softening.  Softening functions are discussed in more detail by \cite{rickaby}   and by  \cite{dorfmann2003,dorfmann}.

None of these models consider the possibility of softening on the primary loading path, i.e. path $A$ of Figure \ref{fig:1}.
\subsection{Softening on the primary loading path}
\cite{mullins1969} observed that in filled rubber vulcanizates pronounced softening occurs during primary loading but only at very small deformations.  He conjectured that this was due to the breakdown of clusters of filler particles. 
For intermediate and large deformations, however, there was no noticeable softening during primary loading.

This property of softening on the primary loading path, i.e. path $A$ of Figure \ref{fig:1}, may  be included within the  model developed here by introducing a new softening function, $\zeta_0^{\phantom{*}}(\lambda)$, which is similar in form to the softening functions (\ref{eq:3.1z}) previously defined, except that $W_{\M}-W$ in (\ref{eq:3.1z}) is replaced by $\lambda_\M-\lambda$.  We therefore define 
$\zeta_0^{\phantom{*}}(\lambda)$ by 
\begin{equation}
\zeta_0^{\phantom{*}}(\lambda)=1-\frac{1}{r_0}\left\{\tanh\left(\frac{\lambda_{\M}-{\lambda}}{ b_0}\right)\right\}^{{1}/{\vartheta_0}}, \quad \for \quad  1 \leq  \lambda\leq\lambda_{\M},
\label{eq:3.3z}
\end{equation}
where $r_0$, $b_0$ and  $\vartheta_0$ are positive constants.    We model the empirical fact that softening on the primary loading path occurs only for small strains by requiring $\zeta_0^{\phantom{*}}(\lambda)$ to be close to 1 when $\lambda$ is close to $\lambda_\M$.

Upon combining $\zeta_0^{\phantom{*}}(\lambda)$ with equation (\ref{eq:2.7z}) for an incompressible, isotropic material the initial primary loading path can be modelled by,
\begin{equation}
\textbf{T}^{\mathscr{E}_{\iso}}(\lambda)=\zeta_0^{\phantom{*}}(\lambda)\left\{-p\textbf{I}+\left[\frac{\mu}{3} \sqrt{\frac{3N} {I_1}}\mathscr{L}^{-1}\left(\sqrt{\frac{I_1}{3{ N }}}\right)\right]\textbf{B}\right\}.
\label{eq:3.4z}
\end{equation}
Eliminating the pressure $p$ by the requirement that $\textbf{T}^{\mathscr{E}_{\iso}}_{22}(\lambda)=\textbf{T}^{\mathscr{E}_{\iso}}_{33}(\lambda)=0$, as before, gives the uniaxial tension
\begin{align}
T^{\mathscr{E}_{\iso}}_{11}(\lambda) &=  \frac{2\mu}{3}\zeta_0^{\phantom{*}}(\lambda)(\lambda^2-\lambda^{-1})\sqrt{\frac{3 N}{I_1}}\mathscr{L}^{-1}\left(\sqrt{\frac{I_1}{3N }}\right).
\label{eq:3.5z}
\end{align}

As $\zeta_0^{\phantom{*}}(\lambda)$ has been modelled to be very close to $1$ on the primary loading path when $\lambda$ is close to $\lambda_\M$, it will not affect the modelling of any subsequent primary loading, or    unloading and reloading, paths.

The inclusion of a softening function on the primary loading path has not previously been discussed in the literature in relation to the Mullins effect.

\section{Transversely isotropic elastic response} 
\label{sec:anisotropic}
 \cite{dorfmann} found that after a uniaxial deformation such as (\ref{eq:2.1z}), there is  an induced change in the material symmetry because some of the damage caused by the stretch is irreversible.   The material symmetry therefore changes from being fully isotropic to being transversely isotropic with preferred direction in the direction of uniaxial stretch.   This change of material symmetry  influences all of the  subsequent response of the material.
\cite{horgan} conjectured that if loading is terminated at the stretch $\lambda_{\M}$ on the primary loading path, then the damage caused is  dependent on the value of $\lambda_{\M}$ and that this must be reflected in the subsequent response of the material upon unloading and  reloading. They too concluded that the material response must become transversely isotropic. 
 
   \cite{diani2005} have also observed experimentally
the transition from an isotropic material to an anisotropic one for carbon filled elastomers under uniaxial testing.   
  Strain-induced anisotropy has been studied by other authors, including \cite{park} and \cite{pancheri}.
 
\cite{spencer} characterized a transversely isotropic elastic solid by the existence of a single preferred direction, denoted by the unit vector field $\textbf{A}(\textbf{X})$.  After deformation the preferred direction 
$\textbf{A}(\textbf{X})$ becomes parallel to
 \[\bg{a}=\mathbf{FA}, \]
 which is not in general a unit vector.

The strain energy function $W$ in a transversely isotropic material is a function of five invariants, namely, the three defined by (\ref{eq:2.2z}) and the further two defined by  
\begin{equation}
\begin{split}
I_4&= \tr \{\textbf{C}(\textbf{A}\otimes\textbf{A})\} =   \textbf{A}\cdot(\textbf{C}\textbf{A}) = \bg{a}\cdot\bg{a},  
 \\
I_5&= \tr \{\textbf{C}^2(\textbf{A}\otimes\textbf{A})\} =   \textbf{A}\cdot(\textbf{C}^2\textbf{A}) =\bg{a}\cdot(\mathbf{B}\bg{a}).  
\end{split}
 \label{eq:4.1z}
\end{equation}

For an incompressible material  $I_3 = 1$  and so the strain energy takes the form
\[ W=W(I_1, I_2, I_4, I_5). \] 
The elastic stress in an incompressible transversely isotropic elastic material is then given by
\begin{align}
\textbf{T}^{\mathscr{E}_{\ti}}(\lambda) =\, -p\textbf{I}+2\bigg\{&\left(\frac{\partial{W}}{\partial I_1}+I_1
\frac{\partial{W}}{\partial I_2}\right)
\textbf{B}-\frac{\partial{W}}{\partial I_2}\textbf{B}^2  \nonumber\\
& + \frac{\partial{W}}{\partial
I_4}\bg{a}\otimes\bg{a}+\frac{\partial{W}}{\partial
I_5}\big(\bg{a}\otimes\textbf{B}\bg{a}+\textbf{B}\bg{a}\otimes\bg{a}\big)\bigg\},
\label{eq:4.2z}
\end{align}
where $p$ is an arbitrary pressure and $\otimes$ denotes a dyadic product.  
Equation (\ref{eq:4.2z}) is equivalent to that presented by \cite[eqn (67)]{spencer}.

The preferred direction $\textbf{A}$ lies in the direction of the uniaxial tension (\ref{eq:2.1z}), so that, in components,
\begin{equation}
\textbf{A} = \left(\begin{array}{c}1\\0\\0\end{array}\right),\quad
\bg{a} = \left(\begin{array}{c}\lambda \\0\\0\end{array}\right),\quad
\bg{a}\otimes \bg{a}
= \left( \begin{array}{ccc}	
\lambda^2& 0 & 0\\
0 & 0 & 0\\
0 & 0 & 0 \end{array}\right)
.\\[0pt]
\label{eq:4.3z}
\end{equation}
From equations (\ref{eq:4.1z}) and  (\ref{eq:4.3z})  the invariants $I_4$ and $I_5$ are given by 
\begin{align*}
 I_4&=\lambda^2,\quad I_5=\lambda^4.
 \end{align*}
The stress  (\ref{eq:4.2z}) reduces to
\begin{align*}
\textbf{T}^{\mathscr{E}_{\ti}}(\lambda) =\, -p\textbf{I}+2\bigg\{&\left(\frac{\partial{W}}{\partial I_1}+I_1
\frac{\partial{W}}{\partial I_2}\right)
\textbf{B}-\frac{\partial{W}}{\partial I_2}\textbf{B}^2  \nonumber\\
& + \left(\frac{\partial{W}}{\partial
I_4}+ 2 \lambda^2\frac{\partial{W}}{\partial
I_5}\right)
\bg{a}\otimes\bg{a}\bigg\},
\label{eq:4.2y}
\end{align*}

For the stress to vanish in the reference configuration we require  
\begin{equation}
 \frac{\partial{}}{\partial{I_4}}W(3,3,1,1)=0, \quad  \frac{\partial{}}{\partial{I_5}}W(3,3,1,1)=0. 
 \label{eq:4.4z}
\end{equation}

The invariant $I_4 = \bg{a}\cdot \bg{a}$ is clearly a measure of stretch in the direction of transverse isotropy.  
\cite{merodio} observed that the invariant $I_5$ is more connected with shear stresses acting normally to the preferred direction.
For the uniaxial deformation  (\ref{eq:2.1z})  there are no such shear stresses and so in our elastic model we take the strain energy to be independent of $I_5$.  A strain energy function  that is independent of $I_5$, vanishes in the reference configuration and satisfies the derivative condition  (\ref{eq:4.4z})$_1$  is given by
\begin{equation}
W_{\ti}=\tfrac12 s_1 I_4^{-1}(I_4-1)^2+\tfrac12 s_2  I_4^{-1} (I^{\frac{1}{2}}_4-1)^2,
\label{eq:4.5z}
\end{equation}
where $s_1$ and $s_2$ are constants.
The transversely isotropic strain energy function $W_{\ti}$ above is found in Section \ref{sec:constitutive} to fit the experimental data extremely well.

\section{The  eight-chain model in transverse isotropy} 
\label{sec:transversely}

We follow \cite{kuhl} in developing a model of transversely isotropic elasticity based on the original  
\cite{arruda} eight-chain  model of  isotropic elasticity.   Rubber is regarded as being composed of cross-linked polymer chains, each chain consisting of $N$ links, with  each link being of length $l$.
We introduce the two lengths
\begin{equation}
 r_{\rm L}^{\phantom{L}} = Nl,\qquad r_0 = \sqrt{N}l.
 \label{eq:5.1z}
\end{equation}
The locking length $ r_{\rm L}^{\phantom{L}}$ is the length of the polymer chain when fully extended.  
The chain vector length $r_0$ is the distance between the two ends of the chain in the undeformed configuration. 
Because of significant coiling of the polymer chains this length is considerably less than the locking length.
The value $ r_0 = \sqrt{N}l$ is derived by statistical considerations.

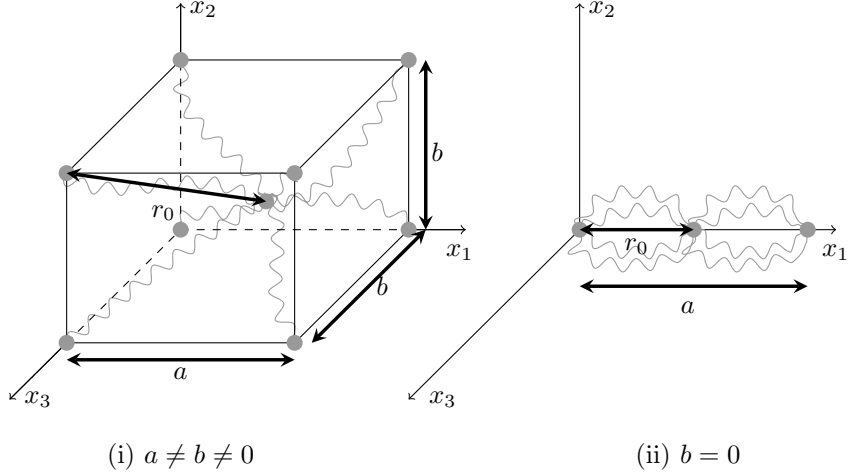
\begin{figure}[h]
\centering
\begin{tikzpicture}[scale=0.75]
%
\draw [black!40,decorate,decoration=snake] 
(1,1) to [out=40,in=210] node [sloped,below]{} (4.5,3.5)
(5,1) to [out=140,in=-60] node [sloped,below]{} (4.5,3.5)
(7,3) to [out=140,in=0] node [sloped,below]{} (4.5,3.5)
(3,3) to [out=65,in=225] node [sloped,below]{} (4.5,3.5)
(4.5,3.5) to [out=100,in=-200] node [sloped,below]{} (5,4)
(4.5,3.5) to [out=140,in=-30] node [sloped,below]{} (1,4)
(3,6) to [out=-120,in=120] node [sloped,below]{} (4.5,3.5)
(4.5,3.5) to [out=10,in=230] node [sloped,below]{} (7,6);
\draw [->](1,1) to [out=225,in=45] node [sloped,below]{} (0,0);
\draw [->](7,3) to[out=0,in=180] node [sloped,above] {} (8,3);
\draw [->](3,6) to[out=90,in=270] node [sloped,above] {} (3,7);
\draw [postaction={decorate}]
(0,0) to [out=45,in=225] node [sloped,below]{} (1,1)
(5,1) to [out=45,in=225] node [sloped,below]{} (7,3)
(1,4) to [out=45,in=225] node [sloped,below]{} (3,6)
(5,4) to [out=45,in=225] node [sloped,below]{} (7,6)
(7,3) to[out=0,in=180] node [sloped,above] {} (8,3)
(1,4) to[out=0,in=180] node [sloped,above] {} (5,4)
(1,1) to[out=0,in=180] node [sloped,above] {} (5,1)
(3,6) to[out=0,in=180] node [sloped,above] {} (7,6)
(1,1) to[out=90,in=270] node [sloped,above] {} (1,4)
(5,1) to[out=90,in=270] node [sloped,above] {} (5,4)
(3,6) to[out=90,in=270] node [sloped,above] {} (3,7)
(7,3) to[out=90,in=270] node [sloped,above] {} (7,6);
\coordinate [label=above:{$x_3$}] (A) at (0.5,-0.3);
\coordinate [label=left:{$x_1$}] (B) at (8.3,2.6);
\coordinate [label=below:{$x_2$}] (C) at (3.4,7.2);
\coordinate [label=above:{$a$}] (D) at (3,0.2);
\coordinate [label=below:{(i) $a\ne b \ne 0$}] (A) at (3,-0.6);
\coordinate [label=left:{$b$}] (E) at (6.8,2);
\coordinate [label=below:{$b$}] (F) at (7.5,4.7);
\draw[dashed][postaction={decorate}]
(1,1) to [out=45,in=225] node [sloped,below]{} (3,3)
(3,3) to[out=0,in=180] node [sloped,above] {} (7,3)
(3,3) to[out=90,in=270] node [sloped,above] {} (3,6);
\fill [black!40] (7,6) circle (4pt);
\fill [black!40] (3,6) circle (4pt);
\fill [black!40] (3,3) circle (4pt);
\fill [black!40] (7,3)circle (4pt);
\fill [black!40] (1,1)circle (4pt);
\fill [black!40] (5,1)circle (4pt);
\fill [black!40] (5,4)circle (4pt);
\fill [black!40] (1,4)circle (4pt);
\fill [black!40] (4.5,3.5)circle (4pt);
\draw [stealth-stealth,line width=1.3pt](4.5,3.5) -- (1,4);
\draw [,stealth-stealth,line width=1.3pt](1,0.7) -- (5,0.7);
\draw [stealth-stealth,line width=1.3pt](5.3,1) -- (7.3,3);
\draw [stealth-stealth,line width=1.3pt](7.3,3) -- (7.3,6);
\coordinate [label=below:{$r_0$}] (A) at (2.7,3.6);
%
%
%
%
\draw[<-](7,0) to [out=45,in=225] node [sloped,below]{} (10,3);
\draw[->](10,3) to[out=0,in=180] node [sloped,above] {} (14.5,3);
\draw[->](10,3) to[out=90,in=270] node [sloped,above] {} (10,7);
\coordinate [label=above:{$x_3$}] (A) at (7.6,-0.3);
\coordinate [label=left:{$x_1$}] (B) at (14.9,2.6);
\coordinate [label=below:{$x_2$}] (C) at (10.4,7.2);
\fill [black!40] (10,3) circle (4pt);
\fill [black!40] (12,3) circle (4pt);
\fill [black!40] (14,3)circle (4pt);
\draw [black!40,decorate,decoration=snake] 
(10,3) to [out=80,in=110] node [sloped,below]{} (12,3);
\draw [black!40,decorate,decoration=snake] 
(10,3) to [out=30,in=150] node [sloped,below]{} (12,3);
\draw [black!40,decorate,decoration=snake] 
(10,3) to [out=240,in=280] node [sloped,below]{} (12,3);
\draw [black!40,decorate,decoration=snake] 
(10,3) to [out=320,in=210] node [sloped,below]{} (12,3);
\draw [black!40,decorate,decoration=snake] 
(12,3) to [out=80,in=110] node [sloped,below]{} (14,3);
\draw [black!40,decorate,decoration=snake] 
(12,3) to [out=30,in=150] node [sloped,below]{} (14,3);
\draw [black!40,decorate,decoration=snake] 
(12,3) to [out=320,in=210] node [sloped,below]{} (14,3);
\draw [black!40,decorate,decoration=snake] 
(12,3) to [out=240,in=280] node [sloped,below]{} (14,3);
\draw [stealth-stealth,line width=1.3pt](10,2) -- (14,2);
\draw [stealth-stealth,line width=1.3pt](10,3) -- (12,3);
\coordinate [label=below:{$a$}] (A) at (11.9,1.9);
\coordinate [label=below:{(ii) $b=0$}] (A) at (11.9,-0.6);
\coordinate [label=below:{$r_0$}] (A) at (11.0,3);
\end{tikzpicture}
\caption{\cite{arruda} eight-chain transversely isotropic model   (i) transversely isotropic case $a\ne b \ne 0$, (ii) special case with $b=0$.}
\label{fig:3}   
\end{figure}

In this extension of the Arruda-Boyce model we consider a cuboid aligned with its edges parallel to the coordinate axes, as in Figure \ref{fig:3}(i).  The edges parallel to the $x_1$-axis, which is the preferred direction of transverse isotropy, have length $a$ and the remaining edges all have length $b$.  Each of the eight vertices of the cuboid is attached to the centre point of the cuboid by a polymer chain, as depicted in Figure \ref{fig:3}(i).  Each of these eight chains is of the same length which we take to be the vector chain length $r_0$.

Using  Figure \ref{fig:3}(i) and equation (\ref{eq:5.1z})$_2$ we see that   the chain vector length may be written
\begin{equation}
r_0=\sqrt N l=\sqrt{\left(\frac{1}{2}a\right)^2+\left(\frac{1}{2}b\right)^2+\left(\frac{1}{2}b\right)^2} =
\frac{a}{2} \sqrt{1+2\alpha^2},
\label{eq:5.2z}
\end{equation}
where $\alpha = b/a$ is the aspect ratio of the cuboid,
 with
$\alpha = 1$ corresponding to material isotropy.
The special case $\alpha=0$  is illustrated in Figure \ref{fig:3}(ii) and  corresponds to $b=0$.  \cite{kuhl} discuss this special case and regard it as representing unidirectional fibre reinforcement.

Now suppose that the material undergoes triaxial extension in directions parallel to the cuboid edges, so that the new dimensions of the cuboid are $(a\lambda_1, b\lambda_2, b\lambda_3)$.  Each of the eight chains has the same length and this new length is given by
\[
r_\chain=\sqrt{\left(\frac{1}{2}a\lambda_1\right)^2+\left(\frac{1}{2}b\lambda_2\right)^2+\left(\frac{1}{2}b\lambda_3\right)^2},
\]
which can be rewritten in terms of the invariants  $I_1$ and $I_4$ as
\begin{equation}
r_\chain=\frac{a}{2}\sqrt{I_4 +\left(I_1-I_4\right)\alpha^2}.
\label{eq:5.3z}
\end{equation}

The argument of the inverse Langevin function is, as in \cite{arruda}, 
\[\frac{r_\chain}{r_{\rm L}^{\phantom{L}}},\]
where $r_{\rm L}^{\phantom{L}}$ is given by equation (\ref{eq:5.1z})$_1$.  We have, using also equations 
(\ref{eq:5.1z})$_2$ and  (\ref{eq:5.2z}),
\[  \frac{r_\chain}{r_{\rm L}^{\phantom{L}}}  = \frac{r_\chain}{r_0}\cdot\frac{r_0}{Nl} = 
\frac{\sqrt{I_4 +\left(I_1-I_4\right)\alpha^2}}{\sqrt{1+2\alpha^2}}\cdot \frac{\sqrt{N}l}{Nl} = 
\sqrt{\frac{I_4 +\left(I_1-I_4\right)\alpha^2}{N(1+2\alpha^2)}}.  \]
The quantities $\gamma$ and $\beta$ are defined by
\begin{align}
\gamma & =  \sqrt{\frac{I_4 +\left(I_1-I_4\right)\alpha^2}{N(1+2\alpha^2)}},   \qquad
\beta  =\mathscr{L}^{-1}\left(\gamma\right),
 \label{eq:5.4z}
\end{align}
where, as before, $\alpha = b/a$ is the aspect ratio of the cuboid in this extension of the Arruda-Boyce model.
We recall that $\alpha = 1$ corresponds to material isotropy so that $I_4$ then cancels out of  equation (\ref{eq:5.4z}) 
reducing it to
\begin{equation}
\gamma = \sqrt\frac{I_1}{3 N },\qquad
\beta =\mathscr{L}^{-1}\left(\gamma\right),
\label{eq:5.5z}
\end{equation}
which is consistent with equation (\ref{eq:2.6z}) of the isotropic Arruda-Boyce model.

By substituting equations (\ref{eq:5.4z}) into the strain energy (\ref{eq:2.5z}) of the isotropic Arruda-Boyce model
 we obtain the following expression for the strain energy in the transversely isotropic Arruda-Boyce model:
\begin{align}
W_{\textrm{A-B}}=&\,\mu N \left\{\gamma \beta+\log \left(\frac{\beta}{\sinh \beta}\right) \right\} -{\frac12}h_4 \left\{I_4-1\right\},
\label{eq:5.6z}
\end{align}
where  $h_4 $ is a constant chosen so that the stress vanishes in the  undeformed state, i.e. chosen so that equation (\ref{eq:4.4z})$_1$ is satisfied.

The stress resulting from equation  (\ref{eq:5.6z}) is
\[ \mathbf{T} = -p\mathbf{I} + 
\mu\frac{\alpha^2}{1+2\alpha^2} \gamma^{-1}\beta\, \mathbf{B} + 
\left(\mu\frac{1 - \alpha^2}{1+2\alpha^2} \gamma^{-1}\beta - h_4 \right) \bg{a}\otimes\bg{a}.
\]
For this stress to vanish in the reference configuration, where  $I_1=3, I_4=1$ and $\gamma = \sqrt{\frac{1}{N}}$, we must take
\[
h_4 =\mu \frac{1-\alpha^2}{1+2\alpha^2} \sqrt N  \mathscr{L}^{-1}\left(\sqrt{\frac{1} N }\right).
\]
For an isotropic material, $\alpha=1$ and we find that $h_4=0$, as expected.

The total strain energy  is obtained by adding  contributions from (\ref{eq:5.6z}) and  (\ref{eq:4.5z}):
\[ W= W_{\textrm{A-B}} + W_{\ti}. \]
From (\ref{eq:4.2z}), the total elastic stress in the transversely isotropic Arruda-Boyce  model~is 
\begin{align}
 \mathbf{T}^{\mathscr{E}_{\ti}}(\lambda) = -p\mathbf{I} + &
 \mu\frac{\alpha^2}{1+2\alpha^2} \gamma^{-1}\beta\, \mathbf{B} + 
\left(\mu\frac{1 - \alpha^2}{1+2\alpha^2} \gamma^{-1}\beta - h_4 \right) \bg{a}\otimes\bg{a} \nonumber \\
  &\hspace*{2mm} + I^{-2}_4  \bigg({s_1}(I^{2}_4- 1)+{s_2}(I^{\frac{1}{2}}_4- 1)\bigg)  \bg{a}\otimes\bg{a}.
\label{eq:5.7z}
\end{align}

\section{Stress relaxation in transverse isotropy} 
\label{sec:transverselystress}

\cite{bernstein} developed a model  for non-linear stress relaxation   which has been found to  represent accurately experimental data for stress-relaxation, see 
\cite{tanner} and the references therein.

For a transversely isotropic incompressible viscoelastic solid, we can build on the work of  \cite[pages 114--116] {lockett} and  \cite[Section 12]{Wineman2009} to write down the following version of the \cite{bernstein} model for the relaxation stress
 $\textbf{T}^{\mathscr{R}_{\ti}}$ in transverse isotropy:
\begin{align}
\textbf{T}^{\mathscr{R}_{\ti}}(\lambda,t) & =-p\textbf{I} +\bigg[ {A}_0+\frac{1}{2} \breve{A}_1(t)(I_1-3)-
\breve{A}_2(t)\bigg]{\textbf{B}}+ \breve{A}_2(t){\textbf{B}}^{2}\nonumber\\
&\;\;+\breve{A}_4(t)(I_4-1)\bg{a}\otimes \bg{a}+\breve{A}_5(t)(I_5-1)
\left(\bg{a}\otimes\textbf{B}\bg{a}+ \textbf{B}\bg{a}\otimes \bg{a}\right),
\label{eq:6.1z}
\end{align}
for   $ t>t_0$.  The first line of (\ref{eq:6.1z})  is that derived by  \cite[pages 114--116] {lockett} for full isotropy.

Eliminating the pressure $p$ from  equation  
 (\ref{eq:6.1z}) by the requirement that  $T^{\mathscr{R}_{\ti}}_{22}=T^{\mathscr{R}_{\ti}}_{33}=0$,   gives   the uniaxial tension
\begin{align}
T^{\mathscr{R}_{\ti}}_{11}(\lambda, t) &= (\lambda^2-\lambda^{-1})
\left[{A}_0+\frac{1}{2} \breve{A}_1(t)
(\lambda-1)^2(1+2\lambda^{-1})
+ \breve{A}_2(t)\left(\lambda^2 -1 + \lambda^{-1} \right)\right] \nonumber \\
&\qquad + (\lambda^2-1)\lambda^2\left\{\breve{A}_4(t) +2\breve{A}_5(t) (\lambda^4 + \lambda^2 )\right\}.
\label{eq:6.2z}
\end{align}
with $T_{11}^{\mathscr{R}_{\ti}}(\lambda, t)$ vanishing for   $t\leq t_0$.
In (\ref{eq:6.2z}), ${A}_0$ is a material constant and  $  \breve{A}_l(t)$,  where
$l\in \{1,2, 4, 5\}$, are material functions  which vanish for  $t\leq t_0$ and are continuous for all $t$. 

Figure \ref{fig:1}   represents a cyclically loaded and unloaded rubber specimen with primary loading occurring along  path $P_0^{\phantom{*}}P_1^{\phantom{*}}$, from the point $P_0^{\phantom{*}}$ at time $t_0$ up to the point $P_1^{\phantom{*}}$ where $\lambda=\lambda_\M$,  which is reached at  time $t_1$.   Stress-relaxation then commences at time $t_1$ and follows the   unloading  path $P_1^{\phantom{*}}P_1^{*}$  down to the position $P_1^{*}$ of zero stress, which is reached at time $t^*_1$ and stretch $\lambda_1^{*} < \lambda$.  Stress-relaxation continues at zero stress and decreasing strain along the path $P_1^{*}P_1^{**}$, which point is reached at time $t_1^{**}$ and stretch $\lambda_1^{**}<\lambda_1^{*}$.
On the  reloading  path $P_1^{**}P_2^{\phantom{*}}$ stress-relaxation proceeds until point $P_2$ is reached, at time $t_2$, where once again $\lambda=\lambda_\M$.     This pattern then continues throughout the unloading and reloading process.

For cyclic stress-relaxation  the material functions $\breve{A}_l(t)$  are replaced by
\begin{equation}
A_l(t)=\left\{\begin{array}{llll}
   \breve{A}_l(\phi_0(t))                                    & \textrm{primary loading}, & t_0^{\phantom{*}}\leq t\leq t_1^{\phantom{*}}, &  \textrm{path}\;\; P_0^{\phantom{*}}P_1^{\phantom{*}}\\[2mm]
\breve{A}_l(\phi_1(t))& \textrm{unloading}, & t_1^{\phantom{*}}\leq t\leq t_1^*, &  \textrm{path}\;\; P_1^{\phantom{*}}P_1^*\\[2mm]
\breve{A}_l(\phi_1(t))& \textrm{stress free}, & t^*_1\leq t\leq t_1^{**}, &  \textrm{path}\;\; P_1^*P_1^{**}\\[2mm] 
\breve{A}_l(\phi_2(t))& \textrm{reloading}, & t^{**}_1\leq t\leq t_2^{\phantom{*}}, &  \textrm{path}\;\; P_1^{**}P_2^{\phantom{*}}\\[2mm]
 \;\; \dots&\;\; \dots&\;\; \dots&\;\; \dots
\end{array}\right.
\label{eq:6.3z}
\end{equation}
where  ${A}_l(t)$, $l \in \{1, 2, 4, 5\}$, are continuous material functions which vanish for  $t\leq t_0$. In equation
 (\ref{eq:6.3z}), {$\phi_0$, $\phi_1$ and $\phi_2$ are continuous functions of time. For simplicity, on the unloading paths and on the stress-free paths we  employ the same function $\phi_1$ as the argument for $A_l(t)$.

In the present model we assume that stress relaxation commences from the point of initial loading at time $t_0$.
Stress-relaxation may proceed at different rates in unloading and reloading. This is governed by the functions $\phi_1$ and $\phi_2$ in equation (\ref{eq:6.3z}).
  Separate functions $\phi_1$ and $\phi_2$ are needed for the unloading and reloading phases, respectively, in order to model better the experimental data in
Section \ref{sec:constitutive}.

Employing equation (\ref{eq:6.3z}), from equation (\ref{eq:6.1z}) we can derive the following isotropic and transversely isotropic relaxation stresses, respectively:
\begin{align}
\textbf{T}^{\mathscr{R}_{\iso}}(\lambda,t) & =-p\textbf{I} +\bigg[ {A}_0+\frac{1}{2} {A}_1(t)(I_1-3)-
{A}_2(t)\bigg]{\textbf{B}}+ {A}_2(t){\textbf{B}}^{2}, \label{eq:6.4zz} \\[4pt]
\textbf{T}^{\mathscr{R}_{\ti}}(\lambda,t) & =-p\textbf{I} +\bigg[ {A}_0+\frac{1}{2} {A}_1(t)(I_1-3)-
{A}_2(t)\bigg]{\textbf{B}}+ {A}_2(t){\textbf{B}}^{2}\nonumber\\
&+{A}_4(t)(I_4-1)\bg{a}\otimes \bg{a}+{A}_5(t)(I_5-1)
\left(\bg{a}\otimes\textbf{B}\bg{a}+ \textbf{B}\bg{a}\otimes \bg{a}\right),
\label{eq:6.4z}
\end{align}
for  $ t>t_0$.

The total  stress for a transversely isotropic relaxing material is then given by
\begin{equation}
\mathbf{T}^{\ti}=\left\{\begin{array}{llll}
 \zeta_0^{\phantom{*}}(\lambda)\{\textbf{T}^{\mathscr{E}_{\iso}}(\lambda) + \textbf{T}^{\mathscr{R}_{\iso}}(\lambda, t) \},& \textrm{loading}, & t_0^{\phantom{*}}\leq t\leq t_1^{\phantom{*}}, & \textrm{path}\;\; P_0^{\phantom{*}}P_1^{\phantom{*}}\\[2mm]                 
\zeta_1(\lambda)\{\textbf{T}^{\mathscr{E}_{\ti}}(\lambda) + \textbf{T}^{\mathscr{R}_{\ti}}(\lambda, t) \},& \textrm{unloading}, & t_1^{\phantom{*}}\leq t\leq t_1^{*}, &  \textrm{path}\;\; P_1^{\phantom{*}}P_1^*\\[2mm]
\phantom{\zeta_1(\lambda)\big\{} {\bf 0} & \textrm{stress free}, & t^*_1\leq t\leq t_1^{**}, &  \textrm{path}\;\; P_1^*P_1^{**}\\[2mm] 
\zeta_2(\lambda)\big\{\textbf{T}^{\mathscr{E}_{\ti}}(\lambda) + \textbf{T}^{\mathscr{R}_{\ti}}(\lambda, t) \big\},& \textrm{reloading}, & t^{**}_1\leq t\leq t_2^{\phantom{*}}, &  \textrm{path}\;\; P_1^{**}P_2^{\phantom{*}}\\[2mm]
 \;\; \dots&\;\; \dots&\;\; \dots&
\end{array}\right. 
 \label{eq:6.5z}
 \end{equation}
where $\textbf{T}^{\mathscr{E}_{\iso}}(\lambda)$,  $\textbf{T}^{\mathscr{E}_{\ti}}(\lambda)$, $\textbf{T}^{\mathscr{R}_{\iso}}(\lambda, t)$ and 
$ \textbf{T}^{\mathscr{R}_{\ti}}(\lambda, t)$ are the  stresses (\ref{eq:3.4z}),  (\ref{eq:5.7z}), (\ref{eq:6.4zz}) and (\ref{eq:6.4z}), respectively.

The total stress (\ref{eq:6.5z}) falls to zero in $t>t_0$ and so we must have
$T^{\mathscr{R}_{\ti}}_{11}<0$ for $t>t_0$, implying that   $T^{\mathscr{R}_{\ti}}_{11}<0$ for $\lambda>1$.
Each of the quantities ${A}_0$, $A_1(t)$, $A_2(t)$, $A_4(t)$, $A_5(t)$ occurring in equation (\ref{eq:6.4z}) has positive coefficient for $\lambda>1$ and so at least one of them must be negative to maintain the requirement  $T^{\mathscr{R}_{\ti}}_{11}<0$ for $\lambda>1$.

\section{Creep of residual strain in transverse isotropy} 
\label{sec:residual}

We postulate that the residual strain that is apparent after a loading and unloading cycle   is caused by creep during that cycle and any previous cycles.   The creep of residual strain may proceed at different rates in unloading and reloading.  We consider that the creep of residual strain  does not operate during primary loading.

Following on from the work of \cite{bergstrom}, \cite{rickaby} showed that during cyclic unloading and reloading the creep causing residual strain can be modelled, in the case of isotropy, as a stress  of the form  
\begin{equation}
\textbf{T}^{\mathscr{C}}(\lambda, t)=-p \textbf{I}+\left\{d_\omega\left[\lambda_{\chain}-1\right]^{-1}
\left\{1+\left[\tanhs \breve{a}(t)\right]^{a_1}\right\}  \right\} \textbf{B}, 
\label{eq:7.1z}
\end{equation}
for $t > t_1$ and $\lambda>1$ with $\textbf{T}^{\mathscr{C}}(\lambda, t)$ vanishing for $t\leq t_1$.  Here, $\lambda_\chain=\sqrt{I_1/3}$ in the case of isotropy.    In equation (\ref{eq:7.1z}),   $a_1$ and $d_\omega$ are material constants,  $d_1$  for    unloading and $d_2$  for    reloading, with $d_2\leq d_1$. The function $a(t)$ is defined by
\begin{equation}
a(t)=\left\{\begin{array}{llll}
   0                                      & \textrm{primary loading}, & t_0^{\phantom{*}}\leq t\leq t_1^{\phantom{*}}, &  \textrm{path}\;\; P_0^{\phantom{*}}P_1^{\phantom{*}}\\[2mm]
\breve{a}(\Phi_1(t-t_1))& \textrm{unloading}, & t_1^{\phantom{*}}\leq t\leq t_1^*, &  \textrm{path}\;\; P_1^{\phantom{*}}P_1^*\\[2mm]
\breve{a}(\Phi_1(t-t_1))& \textrm{stress free}, & t^*_1\leq t\leq t_1^{**}, &  \textrm{path}\;\; P_1^*P_1^{**}\\[2mm] 
\breve{a}(\Phi_2(t-t_1))& \textrm{reloading}, & t^{**}_1\leq t\leq t_2^{\phantom{*}}, &  \textrm{path}\;\; P_1^{**}P_2^{\phantom{*}}\\[2mm]
 \;\; \dots&\;\; \dots&\;\; \dots&\;\; \dots
\end{array}\right.
\label{eq:7.2z}
\end{equation}
where $\Phi_1$ and $\Phi_2$ are continuous functions of time for the unloading and reloading phases, respectively.
For simplicity,  on the stress-free paths we also employ $\Phi_1$ as the argument for  $a(t)$.

The polymer chain length extension ratio is denoted by $\lambda_\chain$ and defined by
\begin{equation}
\lambda_\chain =  \frac{r_\chain}{r_0} =   \sqrt{\frac{I_4 +\left(I_1-I_4\right)\alpha^2}{1+2\alpha^2}} = \sqrt{N}\gamma,
\label{eq:7.3z}
\end{equation}
in which equations (\ref{eq:5.2z})--(\ref{eq:5.4z}) have been used.  In the isotropic case, $\alpha=1$, equation (\ref{eq:7.3z}) reduces to $\lambda_\chain=\sqrt{I_1/3}$ as already recorded after equation (\ref{eq:7.1z}).

The transversely isotropic residual strain model is obtain by substituting $\lambda_{\chain}$  from equation (\ref{eq:7.3z}) into equation (\ref{eq:7.1z}) to give the creep stress
\begin{equation}
\textbf{T}^{\mathscr{C}_{\ti}}(\lambda, t)=-p \textbf{I}+\bigg\{d_\omega\left[\sqrt{N}\gamma-1\right]^{-1}
\left\{1+\left[\tanhs a(t)\right]^{a_1}\right\} \bigg\} \mathbf{B}, 
\label{eq:7.4z}
\end{equation}
for $t > t_1$ and $\lambda>1$.

The total  stress for a transversely isotropic relaxing material is then given by
\begin{equation}
\mathbf{T}^{\ti}=\left\{\begin{array}{llll}
\zeta_0^{\phantom{*}}(\lambda) \textbf{T}^{\mathscr{E}_{\iso}+\mathscr{R}_{\iso}}(\lambda,t),& \textrm{primary loading}, & t_0^{\phantom{*}}\leq t\leq t_1^{\phantom{*}}, & \textrm{path}\;\; P_0^{\phantom{*}}P_1^{\phantom{*}}\\[2mm]                 
\zeta_1(\lambda)\textbf{T}^{\mathscr{E}_{\ti} + \mathscr{R}_{\ti} + \mathscr{C}_{\ti} }(\lambda, t),& \textrm{unloading}, & t_1^{\phantom{*}}\leq t\leq t_1^{*}, &  \textrm{path}\;\; P_1^{\phantom{*}}P_1^*\\[2mm]
\phantom{\zeta_1(\lambda)}{\bf 0} & \textrm{stress free}, & t^*_1\leq t\leq t_1^{**}, &  \textrm{path}\;\; P_1^*P_1^{**}\\[2mm] 
\zeta_2(\lambda)\textbf{T}^{\mathscr{E}_{\ti} + \mathscr{R}_{\ti} + \mathscr{C}_{\ti} }(\lambda, t),& \textrm{reloading}, & t^{**}_1\leq t\leq t_2^{\phantom{*}}, &  \textrm{path}\;\; P_1^{**}P_2^{\phantom{*}}\\[2mm]
 \;\; \dots&\;\; \dots&\;\; \dots&\;\; \dots
\end{array}\right.
 \label{eq:7.5z}
 \end{equation}
 in which for notational convenience we have defined stresses
\begin{align*}
 \textbf{T}^{\mathscr{E}_{\iso} + \mathscr{R}_{\iso}}(\lambda, t)  &= \textbf{T}^{\mathscr{E}_{\iso}}(\lambda) + \textbf{T}^{\mathscr{R}_{\iso}}(\lambda, t),\\[2mm]
  \textbf{T}^{\mathscr{E}_{\ti} + \mathscr{R}_{\ti} + \mathscr{C}_{\ti} }(\lambda, t)  &= \textbf{T}^{\mathscr{E}_{\ti}}(\lambda) + \textbf{T}^{\mathscr{R}_{\ti}}(\lambda, t) + \textbf{T}^{\mathscr{C}_{\ti}}(\lambda, t),
\end{align*} 
where $\textbf{T}^{\mathscr{E}_{\iso}}(\lambda)$,  $\textbf{T}^{\mathscr{E}_{\ti}}(\lambda)$, $ \textbf{T}^{\mathscr{R}_{\iso}}(\lambda, t)$, $ \textbf{T}^{\mathscr{R}_{\ti}}(\lambda, t)$ and $ \textbf{T}^{\mathscr{C}_{\ti}}(\lambda, t)$ are given by equations (\ref{eq:3.4z}),  (\ref{eq:5.7z}), 
 (\ref{eq:6.4zz}), (\ref{eq:6.4z}) and (\ref{eq:7.4z}), respectively.

\section{Constitutive model and comparison with experiment} 
\label{sec:constitutive}
\subsection{Constitutive model}
On substituting the individual stresses given by equations (\ref{eq:5.7z}), (\ref{eq:6.4z}) and (\ref{eq:7.4z}) into equation (\ref{eq:7.5z}) we obtain the following constitutive model for the transversely isotropic material,
\begin{align}
\textbf{T}^{\ti} =&\,\left[1-\frac{1}{r_\omega}\left\{\tanh\left(\frac{W_{\M}-{W}}{\mu b_\omega}\right)\right\}^{{1}/{\vartheta_\omega}}\right]\times\nonumber\\
&\times\Bigg\{ -p\mathbf{I} + 
 \mu\frac{\alpha^2}{1+2\alpha^2} \gamma^{-1}\beta\,  \mathbf{B} + 
\left(\mu\frac{1 - \alpha^2}{1+2\alpha^2} \gamma^{-1}\beta - h_4 \right) \bg{a}\otimes\bg{a}\nonumber\\
&\qquad+\Bigg[ {A}_0+\frac{1}{2} {A}_1(t)(I_1-3)-
{A}_2(t)\Bigg]{\textbf{B}}+ {A}_2(t){\textbf{B}}^{2}\nonumber\\
&\qquad+{A}_4(t)(I_4-1)\bg{a}\otimes \bg{a}+{A}_5(t)(I_5-1)
\left(\bg{a}\otimes\textbf{B}\bg{a}+ \textbf{B}\bg{a}\otimes \bg{a}\right)
\nonumber\\
&\qquad+d_\omega\left[\sqrt{N}\gamma-1\right]^{-1} \left\{1+\left[\tanhs a(t)\right]^{a_1}\right\} \mathbf{B}
\nonumber\\
&\qquad+ I^{-2}_4  \bigg({s_1}(I^{2}_4- 1)+{s_2}(I^{\frac{1}{2}}_4- 1)\bigg)  \bg{a}\otimes\bg{a} \Bigg\} .
\label{eq:8.1z}
\end{align}
\subsection{Comparison with  experimental data}
\label{sec:experimentaldata}
In modelling the Mullins effect we have used the Biot stress $\textbf{T}_B$, defined by
\[
\textbf{T}_B=\lambda^{-1}\textbf{T},
\]
in order to compare our theoretical results with experiment.

Figures \ref{fig:5}   and \ref{fig:6}  provide a comparison of the constitutive model we have developed with experimental data, which came courtesy of \cite{dorfmann} and was presented in their paper.

In comparing our model with experimental data we have employed the Heaviside step function $H(t)$ defined by
\[
H(t)=\left\{ \begin{array}{clrr}
0& t<0,\\
1& t\geq 0,
\end{array}\right.
\]
and we have approximated the inverse Langevin function by its Pad\'e approximant  derived by \cite{cohen}, namely,
\begin{equation}
\mathscr{L}^{-1}(x)\approx\frac{3x-x^3}{1-x^2}.
\label{eq:8.2z}
\end{equation}
This is a very good approximation, even close to the singularity at $x=1$, see \cite[Figure 8]{rickaby}. 
The close relation of the simple model of \cite{gent} to the approximate equation (\ref{eq:8.2z})  has been made clear by \cite{horgan2002}.

Figure \ref{fig:5}   has been obtained by applying the following constants and functions,
\[
N=5.7, \quad \mu=0.56, \quad r_1=r_2=1.67, \quad \alpha^2=0.8, \quad a_1=0.4,
\]
\[
{A}_0=-0.001, \quad {A_{1,2}}(t)= -0.022\log(0.8t), \quad {A_{4,5}}(t)= 0, \quad a(t)=H(t-t_1)t, 
\]
\[
 \zeta_0^{\phantom{*}}(\lambda)=1-0.143\left[\tanhs(3-\lambda)\right]^{1.5},
\]
\[
d_\omega=\left\{ \begin{array}{clrr}
0.0017\\
0.0009\\
\end{array}\right.
\mu b_\omega=\left\{ \begin{array}{clrr}
1.50\\
3.10\\
\end{array}\right.
\vartheta_\omega=\left\{\begin{array}{clrr}
0.70\\
1.00\\
\end{array}\right.
s_1=\left\{ \begin{array}{clrr}
0.72\\
0.72\\
\end{array}\right.
s_2=\left\{\begin{array}{clrr}
-0.4& \textrm{unloading},\\
-0.4& \textrm{loading}.\\
\end{array}\right.
\]
We see in Figure \ref{fig:5} that the transversely isotropic model provides a good fit with experimental data and is a significant improvement on the isotropic model of \cite[Figure 15]{rickaby}.
\begin{figure}
\centering
\includegraphics[width=12.5cm,height=10cm, trim= 0 0 0 0]{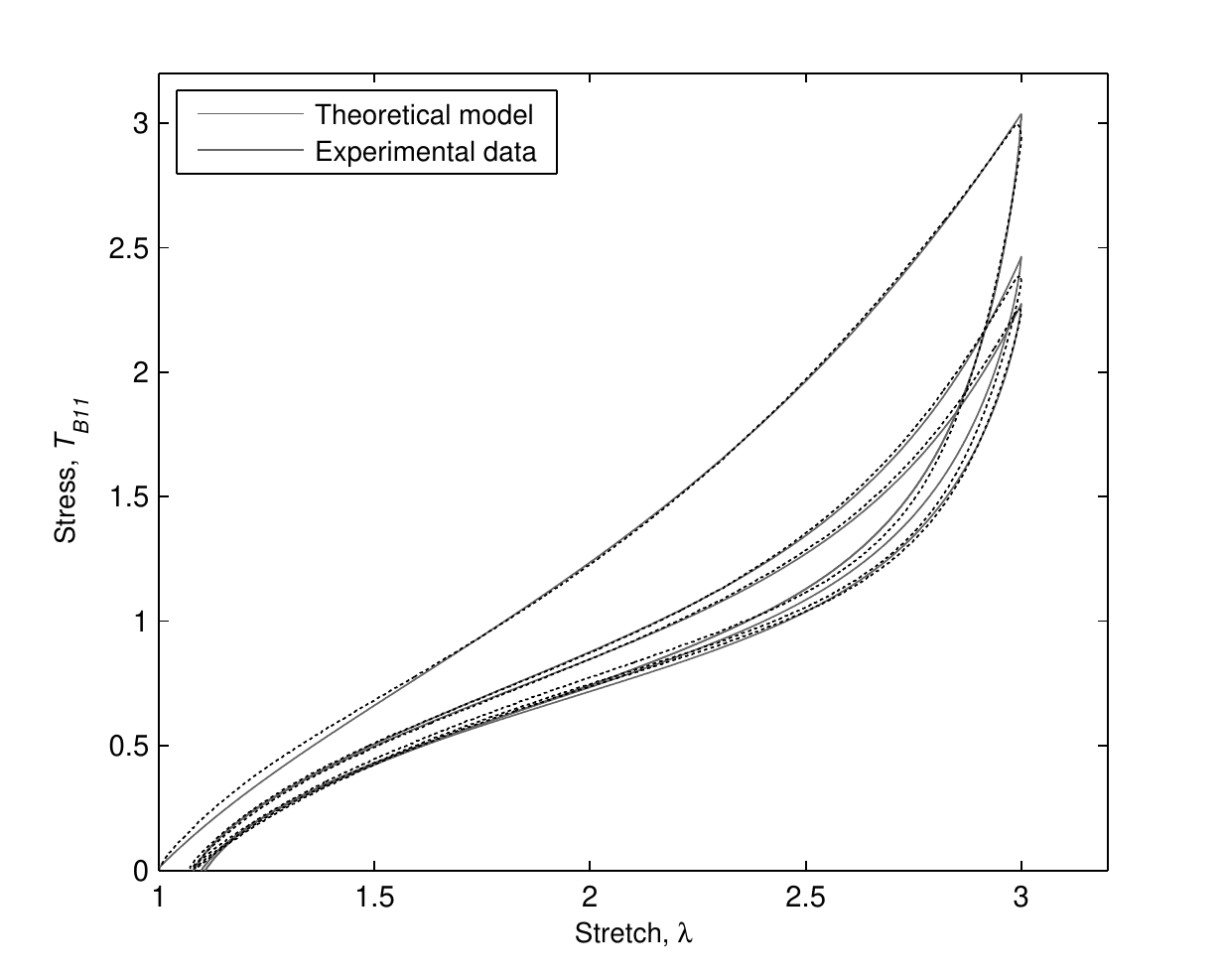}
\caption{Comparison of our theoretical model with experimental data of \cite{dorfmann}, transversely isotropic model, particle-reinforced specimen with 20 phr of carbon black.}
\label{fig:5}   
\end{figure}
Figure \ref{fig:6}   has been obtained by applying the following constants and functions,
\[
N=5.5, \quad \mu=1.41, \quad r_1=r_2=1.14, \quad \alpha^2=0.8, \quad a_1=0.4,
\]
\[
{A}_0=-0.01, \quad {A_{1,2}}(t)= -0.067\log(t), \quad {A_{4,5}}(t)= 0, \quad  a(t)=H(t-t_1)t,
\]
\[
\zeta_0^{\phantom{*}}(\lambda)=1- 0.840\left[\tanhs(3-\lambda)\right]^{3.2},
\]
\[
d_\omega=\left\{ \begin{array}{clrr}
0.020\\
0.012\\
\end{array}\right.
\mu b_\omega=\left\{ \begin{array}{clrr}
1.14\\
2.48\\
\end{array}\right.
\vartheta_\omega=\left\{\begin{array}{clrr}
0.70\\
1.00\\
\end{array}\right.
s_1=\left\{ \begin{array}{clrr}
2.60\\
2.60\\
\end{array}\right.
s_2=\left\{\begin{array}{clrr}
-0.2& \textrm{unloading},\\
-0.2& \textrm{loading}.\\
\end{array}\right.
\]
Figure \ref{fig:6} shows that the transversely isotropic model provides a good fit with experimental data and is a significant improvement on the isotropic model of \cite[Figure 16]{rickaby}.  The better fit here is partly due to the modelling of stress-softening on the primary loading path.

\begin{figure}[h]
\centering
\includegraphics[width=12.5cm,height=10cm, trim= 0 0 0 0]{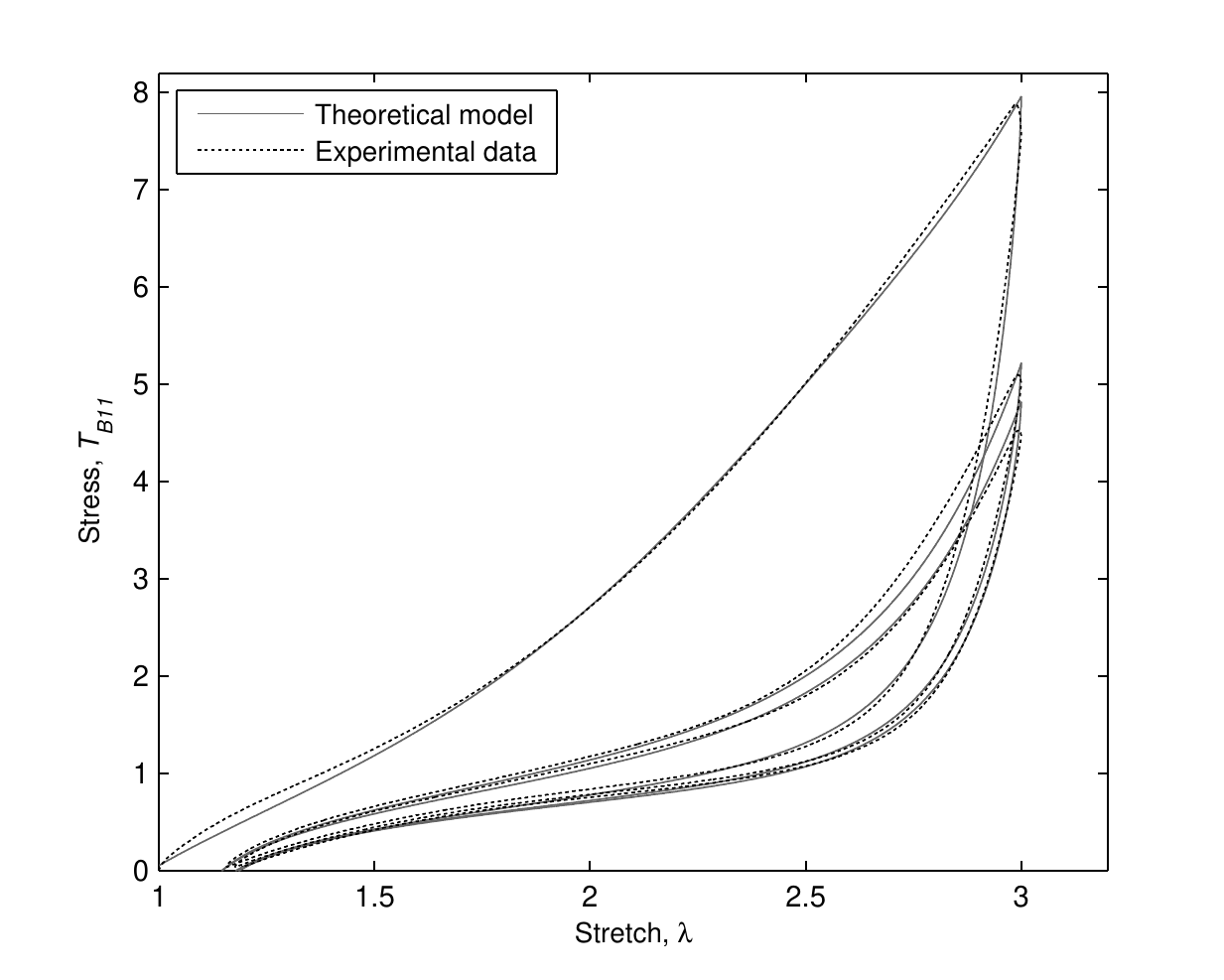}
\caption{Comparison of our theoretical model with experimental data of \cite{dorfmann}, transversely isotropic model, particle-reinforced specimen with 60 phr of carbon black.}
\label{fig:6}   
\end{figure}

\section{Conclusions} 
\label{sec:conclusion}
This model appears to be the first appearance in the literature of
 a transversely isotropic stress-softening and residual strain model which has been combined with  a transversely isotropic version of the Arruda-Boyce eight-chain constitutive model of elasticity in order to develop a model that is capable of accurately representing the Mullins effect in uniaxial tension when compared with  experimental data.

The model has been developed is such a way that any of the salient inelastic features could be excluded and the integrity of the model would still be maintained. The proposed model should prove extremely effective in the modelling of many practical applications.

Figures \ref{fig:5} and \ref{fig:6}   provide a comparison between experimental the data of \cite{dorfmann} and the transversely isotropic model presented here.   
By comparing Figures \ref{fig:5} and \cite[Figure 15]{rickaby} it can be seen that the present transversely isotropic model provides a much better fit to the data than does the original isotropic model of \cite{rickaby}.
Similarly,  for the higher concentration of carbon black of Figures \ref{fig:6} comparing with \cite[Figure 16]{rickaby} shows that the  transversely isotropic model provides an equally better fit.  This is partly due to the new feature of the  modelling of stress-softening on the primary loading path.

After an applied uniaxial deformation, the induced transverse isotropy means that the directions perpendicular and parallel to the deformation  have sustained different degrees of damage which  is borne out by the experimental data of \cite[Figure 3]{diani2005}. Thus, whilst some of the  material parameters of our  model  remain the same along the direction of uniaxial tension and perpendicular to it, the anisotropic terms, in particular the stress-relaxation terms, may not necessarily be equal for two perpendicular directions.

Our present model can be modified to cope with multiple stress/strain cycles, with increasing values of maximum stretch, and we hope to present a comparison between theory and experiment at a later date.

The version of the model developed here is  for uniaxial tension.
We expect that the results presented here could be  extended to include equibiaxial tension, pure or simple shear, and  a general three-dimensional model. These ideas will be developed in later papers.

A further application of the model could be to the mechanics of soft biological tissue. The comparison between the stress-softening associated with soft biological tissue, in particular muscle, and filled vulcanizated rubber has been discussed in detail by \cite[Section 2]{dorfmann2007}. For cyclic stress softening both materials exhibit stress relaxation, hysteresis, creep and creep of residual strain. Similar observations have been made for arterial material, see for example \cite{holzapfel2000}. After preliminary investigations it is apparent that the model presented here could be extended to include biological soft tissue, though  the inherent anisotropy of biological tissue would need to be taken into consideration.

\section*{Acknowledgements}
One of us (SRR) is grateful to  the University of East Anglia for the award of a PhD studentship. The authors thank  Professor Luis Dorfmann for most kindly supplying experimental data.  Furthermore, we would like to thank the reviewers for their constructive comments and suggestions.

\bibliographystyle{natbib}
\bibliography{BIBLIOGRAPHYa}


\end{document}